\documentstyle[twoside,fleqn,espcrc2,psfig]{article}

\def\eq#1{{eq. (\ref{#1})}}

\def\lsim{\raise0.3ex\hbox{$\;<$\kern-0.75em\raise-1.1ex\hbox{$\sim\;$}}}
\def\gsim{\raise0.3ex\hbox{$\;>$\kern-0.75em\raise-1.1ex\hbox{$\sim\;$}}}

\newcommand{\bea}{\begin{eqnarray}}
\newcommand{\eea}{\end{eqnarray}}

\def\beq{\begin{equation}}
\def\eeq{\end{equation}}

\def\bea{\begin{eqnarray}}
\def\ba{\begin{array}}
\def\ea{\end{array}}

\def\ra{\rightarrow}

\def\ib#1#2#3{           {\it ibid. }{\bf #1} (19#2) #3}

\def\pr#1#2#3{           {\it Phys. Rev. }{\bf #1} (19#2) #3}

\def\sjnp#1#2#3{         {\it Sov. J. Nucl. Phys. }{\bf #1} (19#2) #3}

\def\lsim{\;
\raise0.3ex\hbox{$<$\kern-0.75em\raise-1.1ex\hbox{$\sim$}}\;
}

\newcommand{\AmS}{{\protect\the\textfont2
  A\kern-.1667em\lower.5ex\hbox{M}\kern-.125emS}}

\title{
\begin{flushright}
\vglue -2.5cm
{\small
FTUV/96-12  \\
IFIC/96-13  \\
hep-ph/9602384 \\
}
\vglue 0.5cm
\end{flushright}
The MSW solution to the solar neutrino problem 
in the presence of random solar matter density perturbations}

\author{A. Rossi\address{Instituto de F\'{\i}sica Corpuscular - C.S.I.C.\\
Departament de F\'{\i}sica Te\`orica, Universitat de Val\`encia\\
46100 Burjassot, Val\`encia, SPAIN} 
\thanks{Supported by the grant N. ERBCHBI CT-941592 of the Human Capital 
and Mobility Program. This contribution is based on the paper [1]  done in 
collaboration with H. Nunokawa, V. Semikoz and J. W. F. Valle. 
}}
\begin{document}
\renewcommand{\thefootnote}{\fnsymbol{footnote}}

\begin{abstract}
We present the  evolution equation describing  
MSW conversion,  derived in the framework of the Schr\"odinger 
approach, in the presence of  matter density fluctuations.  
Then we analyse the effect of such fluctuations in the 
MSW scenario as a solution to the solar neutrino problem. 
It is shown that the non-adiabatic MSW 
parameter region is rather stable (especially in $\delta m^2$) 
for matter density noise at the few percent level. 
We also discuss the possibility to probe solar matter density 
fluctuations at the future Borexino experiment. 
\end{abstract}
\maketitle
\renewcommand{\thefootnote}{\arabic{footnote}}
\setcounter{footnote}{0}

{\bf 1.}
The present deficit of solar neutrinos seems to disfavour any 
"astrophysical solutions" \cite{astro}
whereas it points to neutrino oscillations. 
In particular the resonant conversion due to 
neutrino interactions with constituents of the solar material  
(the Mikheyev-Smirnov-Wolfenstein  (MSW) effect) \cite{MSW} is the most 
elegant and viable explanation for the existing solar neutrino data. It
provides an extremely good data fit in the small mixing region 
with  $\delta m^2 \simeq 10^{-5}$eV$^2$ and $\sin^2 2 \theta \simeq 10^{-3} 
\div 10^{-2}$ \cite{FIT,smirnov,hata}. 

Here we investigate the stability of the 
MSW solution with respect to the possible presence 
of random perturbations in the solar matter density.

The existence of matter density perturbations at 
the level of 1\% or so cannot be excluded either by 
the Standard Solar Model (SSM), which is based on hydrostatic 
evolution equations,  or by the present helioseismology 
observations  \cite{dal}. 

Let us remind that 
in Ref.\cite{KS} the effect of periodic matter density 
perturbations added to an average density $\rho_0$, i.e.
$\rho(r) = \rho_0  [1 + h \sin (\gamma r)]$
upon resonant neutrino conversion was investigated. In that case 
parametric resonance in the neutrino conversion can occur 
 when the fixed frequency 
($\gamma$) of the perturbation is close to the neutrino 
oscillation eigen-frequency and for rather large 
amplitude ($h \sim 0.1-0.2$).  
There are also a number of papers 
which address similar effects by different 
approaches \cite{AbadaPetcov,BalantekinLoreti}. 

In the present discussion we  consider "white noise" matter dentity 
perturbations $\delta \rho$ (as in Ref. \cite{BalantekinLoreti}).    
Namely we assume that the random field $\delta \rho$ has 
 a $\delta$-correlated Gaussian distribution:
\beq
\label{correlator}
\langle \delta \rho(r_1)\delta \rho(r_2)\rangle = 2\rho^2\langle 
\xi^2\rangle L_0 \delta (r_1 - r_2)
\eeq
where $\xi= \delta \rho/\rho$  and 
the correlation length $L_0$ obeys the following relation:
\beq
\label{size}
l_{{free}} \ll L_0 \ll \lambda_m
\eeq
In (\ref{size}) the lower bound is dictated by the hydro-dynamical 
approximation 
used later on, $l_{\rm free}= (\sigma n)^{-1}$ being the 
mean free path of the particles in the solar matter
\footnote{For Coulomb interactions, the
cross-section $\sigma$  is determined by the classical radius of 
electron $r_{0e} =e^2/m_ec^2\sim 2\times
10^{-13}$cm, 
resulting in 
$l_{\rm free}\sim 10$cm 
for a solar mean density $n_0 \sim 10^{24}$cm$^{-3}$.}.  
On the other hand, the upper bound expresses the fact 
that the scale of fluctuations have to be much smaller 
than the characteristic $\nu$ matter 
oscillation length $\lambda_m$, as indeed the 
$\delta$-correlated distribution in \eq{correlator} requires.

\vspace{0.3cm}

{\bf 2.}
According to the standard Schr\"odinger equation approach, 
we derive now the most general  neutrino evolution equation in 
 random matter density. 
The evolution of  a system of two neutrinos 
$\nu_{e}$ and $\nu_{x}$ ($x=\mu$ or $\tau$)\footnote{Here for 
simplicity we consider only the case of  solar $\nu_e$ conversion into 
active state $\nu_{\mu}$ or $\nu_{\tau}$; 
however the discussion can be extended also 
to the case of conversion into a sterile state \cite{NRSV}.} 
in the solar matter is governed by 
\beq
\label{ev1}
i\frac{d}{dt}\left(\begin{array}{c}
\nu_e \\ \nu_x \end{array}\right)\!=\!
\left (\begin{array}{cc} H_{e}  &  H_{e x} \\
            H_{ex} & H_{x} \end{array}\right)
 \left (\begin{array}{c}\nu_e \\ \nu_x \end{array}\right ), 
\eeq
where the entries of the Hamiltonian matrix are
 \begin{eqnarray}
\label{matdef}
 & & H_e=  2[A_{ex}(t) + \tilde{A}_{ex}(t)], ~~~~ H_x=0 , \nonumber \\
& & H_{ex} = \frac{\delta m^2}{4E}\sin2 \theta , \nonumber \\
  & & A_{ex}(t)  =  \frac{1}{2} [V_{ex}(t) 
 - \frac{\delta m^2}{2E} \cos2 \theta] ,  \nonumber \\
& & \tilde{A}_{ex}(t) = \frac{1}{2} V_{ex}(t)\cdot \xi .
\end{eqnarray}
Here $\theta$ is the neutrino mixing angle in vacuum, $\delta m^2$ the 
mass squared difference, $E$ the neutrino energy  
and the matter potential for the $\nu_e \ra \nu_x$ 
conversion reads 
\beq
\label{vex}
V_{ex}(t) = \frac{\sqrt{2} G_F}{m_p} \rho(t) (Y_e) ,
\eeq
where $m_p$ is the nucleon mass and $Y_e$ is the electron number per nucleon. 

The above system  can be rewritten in terms of the following 
equations:
\begin{eqnarray}
\label{sys}
\dot{I}(t)&\!\! = & \!\!H_e(t)R(t) - 2H_{e{x}}(t)(P(t) - 1/2) 
 \nonumber \\
\dot{R}(t)& \!\!= & \!\!- H_e(t)I(t) \nonumber \\
\dot{P}(t)& \!\! = & \!\!2H_{e{x}}(t)I(t) , 
\end{eqnarray} 
where $P \!= \!|\nu_e|^2$ is the $\nu_e$ 
survival probability,  
$R\!=\!{\mbox R}{\mbox e}(\nu_x^{*}\! \nu_e\!)$ and $I\!=\!
{\mbox I}{\mbox m}(\nu_x^{*}\! \nu_e\!)$, with 
the corresponding initial conditions 
$P(t_0) = 1, \,I(t_0)=0, \,R(t_0) = 0$. 

The Eqs. (\ref{sys}) have to be averaged (see \cite{NRSV} for more details)
over the random density distribution, taking into account that 
for the random component we  have:
\beq
\label{matcorrel}
\langle \tilde{A}_{ex}^{2n+1}\rangle \! = \!0, ~~~
\langle \tilde{A}_{ex}(t)\tilde{A}_{ex}(t_{1})\rangle = 
2\kappa\delta (t - t_{1}) ,
\eeq
where the  quantity $\kappa$ is given by:
\beq
\label{den_noise}
\kappa(t)=  \langle \tilde{A}_{ex}^2(t)\rangle L_0 = \frac{1}{4} 
V^2_{ex}(t)
\langle \xi^2\rangle L_0 .
\eeq
In terms of the averaged quantities defined 
as $\langle P(t)\rangle = \cal {P}(t)$, 
$\langle R(t)\rangle = \cal {R}(t)$, $\langle I(t)\rangle = \cal {I}(t)$, 
we can write the variant of the set (\ref{sys}) as:
\begin{eqnarray}
\label{sys1}
\dot{\cal{I}}(t) \!\! &\! = \!& \!\!\! 
2[ A_{ex}(t) \cal{R}(t) - \kappa(t)\cal{I}(t) 
- H_{ex} (\cal{P}(t)-1/2)]  \nonumber \\
\dot{\cal{R}}(t) \!&\! = \!&\!\! -2[ A_{ex}(t) \cal{I}(t) + 
\kappa(t)\cal{R}(t) ]
 \nonumber \\
\dot{\cal{P}}(t) \!&\!= \!&\!\! \!2 H_{ex} \cal{I}(t). 
\end{eqnarray}
This system of equations\footnote{These equations are equivalent to 
those obtained in Ref.\cite{BalantekinLoreti} in terms of the variables 
$x=2\cal{R}$,  $y=-2\cal{I}$ and $r=2\cal{P}-1$.} 
explicitly exhibits the noise parameter $\kappa$.
It is now possible to outline the main effects due to the 
presence of the random field $\delta \rho$ upon the resonant neutrino 
conversion. The MSW resonance condition remains unaltered, 
i.e. $A_{ex}(t) = V_{ex}(t) -\delta m^2 \cos2\theta/2E = 0$,  
due to the random nature of the matter perturbations. 
Due to the
condition $L_0\ll \lambda_m$, 
the noise parameter 
$\kappa$ (cfr. Eq.(\ref{den_noise})) is always smaller 
than $A_{ex}(t)$ except at the resonance region. 
As a consequence, (see  Eqs. (\ref{sys1})) 
the perturbation can show its 
maximal effect just at the resonance provided that the corresponding noise 
length $1/\kappa$ obeys the following {\it adiabaticity} 
condition at the resonance layer $\Delta r$
\begin{equation}
\label{adiab}
\tilde{\alpha}_r= \Delta r (\kappa)_{res} > 1 .
\end{equation}
This condition is analogous to the standard MSW 
{\it adiabaticity condition} at resonance 
$\alpha_r = \Delta r/(\lambda_m)_{res}>1$ \cite{MSW}. 
For definiteness, we have taken $L_0= 0.1\times \lambda_m$.  
The two adiabaticity parameters are related as 
\beq
\label{alfa}
\tilde{\alpha}_r \approx \alpha_r \frac{\xi^2}{\tan^2 2\theta}, \,\,\,\,\,
\,\,\,\,\,\,\,\,
\alpha_r = \frac{\delta m^2 \sin^2 2 \theta R_0}{4\pi E \cos 2\theta}  .
\eeq
Therefore due to the restriction (\ref{size}) and 
for the  range of parameters we are considering, 
$\xi \sim 10^{-2}$, $\tan^2 2\theta\geq 10^{-3}-10^{-2}$, we have 
$\tilde{\alpha}_r \leq \alpha_r$. 

As a result of (\ref{alfa}), in the adiabatic regime 
$\alpha_r\! >\!1$,  the effect of the noise is enhanced to the extent
that the mixing angle is small. Furthermore, 
the MSW non-adiabaticity $\alpha_r <1$  always brings to 
 $\tilde{\alpha}_r<1$. As a result in our discussion the 
fluctuations are expected to be ineffective in the non-adiabatic 
MSW regime. 
Finally,  it can be shown that the  matter noise weakens the MSW suppression in
 the resonance layer, 
exhibiting somehow the role of a friction. 

\vspace{0.3cm}

{\bf 3.}
In view of the qualitative features just outlined, we discuss the 
implications of noisy solar matter density in the MSW scenario for 
the solar neutrino problem. 
We have solved numerically the coupled differential Eqs. in (\ref{sys1}) 
for the $\nu_e$ survival probability,  
using  as reference SSM the most recent Bahcall-Pinsonneault 
model (BP) \cite{SSM}. 

The $\chi^2$ analysis has been performed taking the latest averaged  
experimental data of chlorine \cite{cl}, gallium \cite{ga,sa} 
and Kamiokande \cite{k} 
experiments:
$R_{Cl}^{exp}= (2.55\pm 0.25) \mbox{SNU}, \,
R_{Ga}^{exp}= (74\pm 8) \mbox{SNU}\footnote{
For gallium result we have taken the weighted average of GALLEX datum 
$R^{exp}_{Ga}= (77\pm8\pm5)$SNU\cite{ga} and SAGE 
$R^{exp}_{Ga}= (69\pm 11\pm 6)$SNU\cite{sa}.}, \,
R_{Ka}^{exp}= (0.44\pm 0.06)R_{Ka}^{BP}$. 
%


\vglue 0.3cm
\hglue -1.0cm
\psfig{file=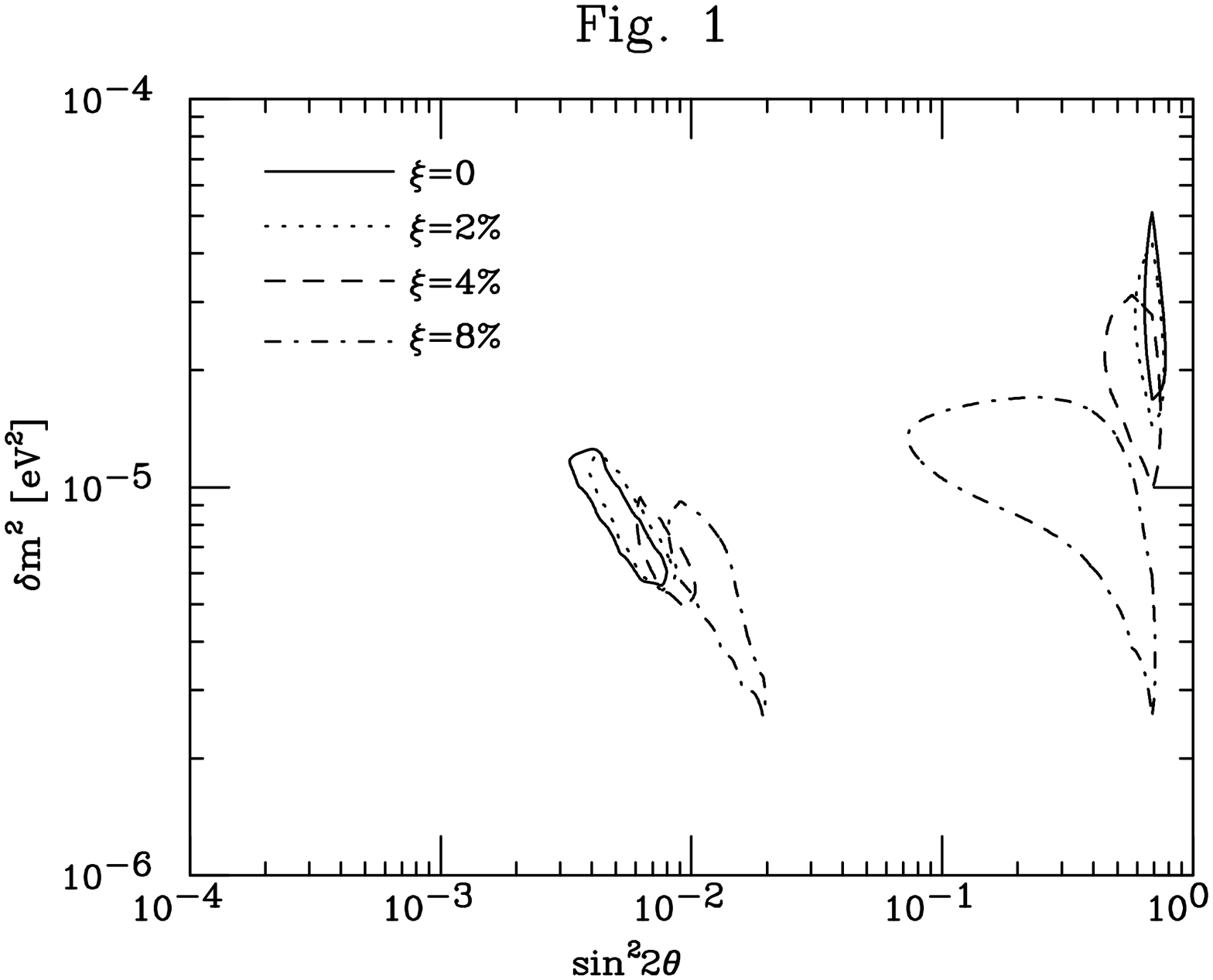,width=9.0cm,height=7.6cm,rwidth=8.0cm,rheight=7.5cm,angle=90}
\vglue -0.1cm

\noindent 
The results of the fitting in the $\delta m^2, \sin^2 2\theta$ parameter 
space, are shown in Fig. 1, 
where the 90\% 
confidence level (C.L.) areas are drawn  for different 
values of $\xi$. 
One can observe that the small-mixing  region is almost stable, 
with a slight shift 
down of $\delta m^2$ values and a slight shift of  
$\sin ^2 2\theta$ towards larger values. 

 The large mixing area is also pretty stable, exhibiting 
the tendency to shift to smaller $\delta m^2$ and $\sin^2 2 \theta$.
The  smaller  $\delta m^2$ values compensate for the 
weakening of the MSW suppression due to the presence of 
matter noise,   so that a larger portion of 
the neutrino energy spectrum can be converted. 
The   $\xi=8\%$ case, considered for the sake of demonstration, 
clearly shows that the small mixing region is 
much more stable than the 
large mixing one even for such large value of the noise. 
Moreover  
the strong selective 
$^7$Be neutrino suppression, which is the nice feature of the MSW effect, 
is somehow degraded by the presence of matter noise.  
Consequently the longstanding conflict between chlorine and Kamiokande data 
is exacerbated and the data fit gets worse. 
Indeed, the presence of the matter 
density noise  makes the data fit a little poorer: 
$\chi^2_{min}= 0.1$  for  $\xi=0$, it 
becomes $\chi^2_{min}= 0.8$ for $\xi=$ 4\% and even 
$\chi^2_{min}= 2$ for $\xi=$8\%. 

In conclusion 
we have shown that the MSW 
solution exists for any realistic levels of matter density noise 
($\xi\leq 4\%$).  
Moreover the MSW solution is essentially stable in mass ($4\cdot 10^{-6}
\mbox{eV}^2 <\delta m^2< 10^{-5}\mbox{eV}^2$ at 90\% CL), whereas 
the mixing appears more sensitive to the level of fluctuations.

\vspace{0.3cm}

{\bf 4.} Let us also stress the fact 
that the solar neutrino experiments could be 
viable tools for  providing 
information on the   matter fluctations in the solar center. 
In particular, the 
future Borexino experiment \cite{borex},  
aiming to detect the $^7$Be neutrino flux   could be 
sensitive to the presence of solar matter fluctuations, as 
the $^7$Be neutrinos are those 
mostly affected by the presence of matter noise.

In the relevant  MSW paramter region for the noiseless case,  
the Borexino signal cannot be definitely predicted 
(see  Fig. 2a). Within the present allowed C.L. regions (dotted line)   
the expected rate,  $Z_{Be}\!=\!R^{pred}_{Be}/R^{SSM}_{Be}$ (solid lines), 
is in the range $0.2\div 0.7$. 

On the other hand, when the  matter density noise is switched on, e.g. 
 $\xi= 4\%$ (see Fig. 2b), the minimal 
allowed value for $Z_{Be}$ becomes higher, $Z_{Be}\!\geq \!0.4$. 
Hence,  if the MSW mechanism is responsable for the 
solar neutrino deficit and Borexino
experiment  detects a low signal, say $Z_{Be}\lsim 0.3$
(with good accuracy)  this will imply that a 4\% level of matter 
fluctuations in the central region of the sun is unlikely . 

\begin{figure}[hbt]
\hglue -0.6cm
\psfig{file=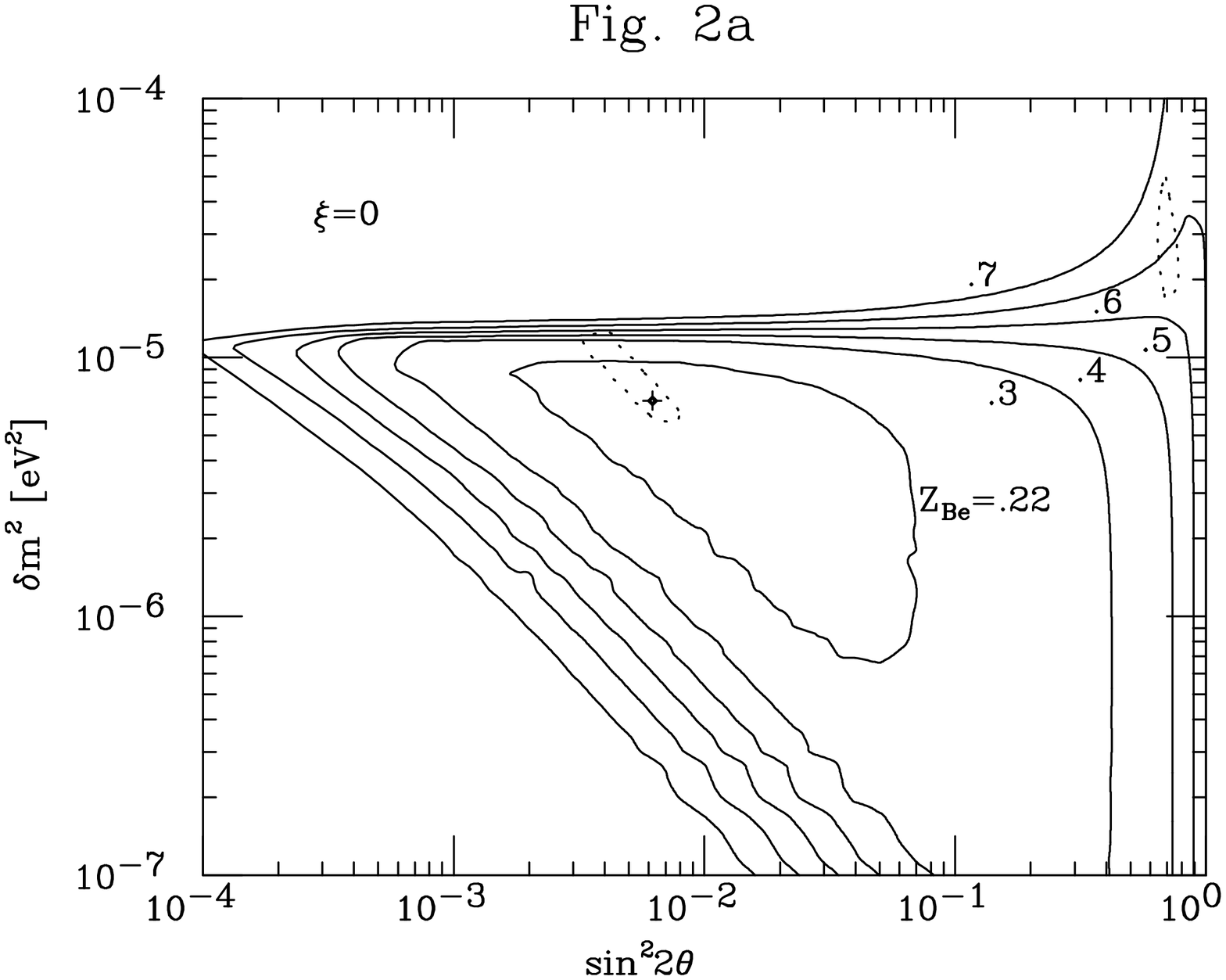,width=9.2cm,height=7.6cm,rwidth=7.5cm,rheight=7.5cm,angle=90}
\vglue -0.4cm
\hskip -0.6cm
\psfig{file=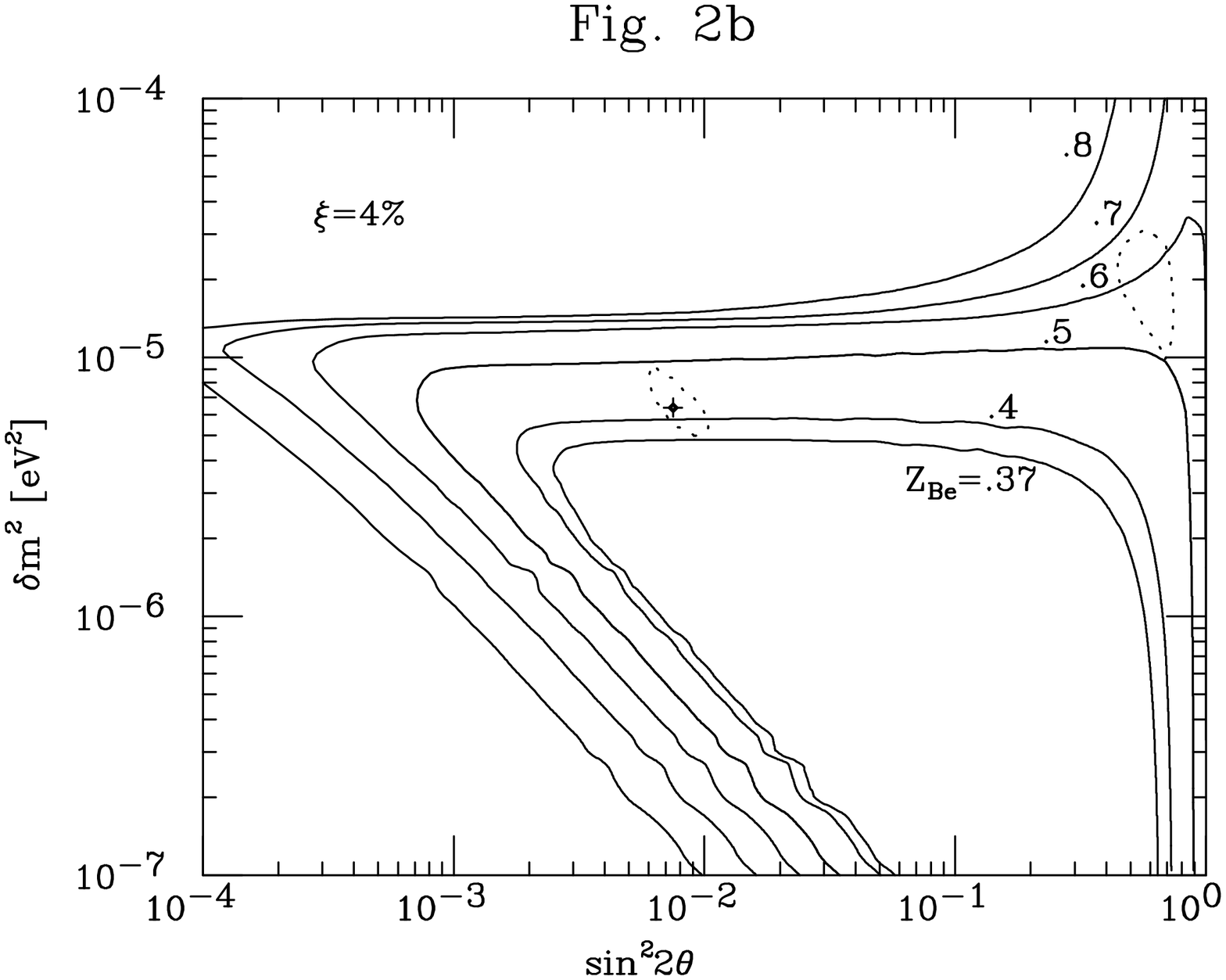,width=9.2cm,height=7.6cm,rwidth=7.5cm,rheight=7.5cm,angle=90}
\vglue -1.0cm
\end{figure}

Once more, the solar neutrino detection turns out to be  
an important approach for studying the solar physics. 

\vspace{0.5cm}

We thank Z. Berezhiani, P. Krastev, S. Mikheyev,  A. Smirnov  and 
S. Turck-Chi$\acute{e}$ze for 
valuable comments and discussion.

\end{document}